\documentclass[pra,superscriptaddress,nofootinbib,twocolumn]{revtex4-2}
\usepackage{amsmath}
\usepackage{amssymb}
\usepackage{amsthm}
\usepackage{bbm}
\usepackage{hyperref}
\usepackage{color}
\usepackage{graphicx}[floatfix]
\usepackage{hyperref}
\usepackage{physics}
\usepackage[T1]{fontenc}
\usepackage{times}
\usepackage[utf8]{inputenc}

\hypersetup{
    colorlinks=true,linkcolor=blue,citecolor=blue,
    filecolor=blue,urlcolor=blue,breaklinks=true
  }

\newcommand{\e}[1]{\{#1\}}

\newcommand{\orcid}[1]{\href{https://orcid.org/#1}{\includegraphics[width=7pt]{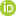}}}

\theoremstyle{definition}

\def\be{\begin{equation}}
\def\ee{\end{equation}}

\def\bc{\begin{center}}
\def\ec{\end{center}}
\def\bal{\begin{align}}
\def\eal{\end{align}}

\begin{document}
\title{Quantum graph models for transport in filamentary switching}
\author{Alison A. Silva\orcid{0000-0003-3552-8780}}
\affiliation{
  Programa de Pós-Graduação em Ciências/Física,
  Universidade Estadual de Ponta Grossa,
  84030-900 Ponta Grossa, Paraná, Brazil
}


\author{Fabiano M. Andrade\orcid{0000-0001-5383-6168}}
\affiliation{
  Programa de Pós-Graduação em Ciências/Física,
  Universidade Estadual de Ponta Grossa,
  84030-900 Ponta Grossa, Paraná, Brazil
}
\affiliation{
  Departamento de Matemática e Estatística,
  Universidade Estadual de Ponta Grossa,
  84030-900 Ponta Grossa, Paraná, Brazil
}

\author{Francesco Caravelli\orcid{0000-0001-7964-3030}}
\affiliation{
 Theoretical Division (T4),
Los Alamos National Laboratory, Los Alamos, New Mexico 87545, USA
}

\date{\today}

 \begin{abstract}
The formation of metallic nanofilaments bridging two electrodes across an insulator is a mechanism for resistive switching. Examples of such phenomena include atomic synapses, which constitute a distinct class of memristive devices whose behavior is closely tied to the properties of the filament. Until recently, experimental investigation of the low-temperature regime and quantum transport effects has been limited. However, with growing interest in understanding the true impacts of the filament on device conductance, comprehending quantum effects has become crucial for quantum neuromorphic hardware.

We discuss quantum transport resulting from filamentary switching in a narrow region where the continuous approximation of the contact is not valid, and only a few atoms are involved. In this scenario, the filament can be represented by a graph depicting the adjacency of atoms and the overlap between atomic orbitals. Using the theory of quantum graphs with node scattering which is locally diffusive, we calculate the scattering amplitude of charge carriers on this graph and explore the interplay between filamentary formation and quantum transport effects.
\end{abstract}

\maketitle


\section{Introduction}

There has been a huge interest in resistive switching and memristive behavior, as the complexity of many nanoscale devices requires new techniques to understand their behavior.

For instance, memristors offer the possibility of harnessing both nonlinear behavior and non-trivial memory in electronic circuits. Specialized circuits composed of large numbers of such devices promise a new generation of computational hardware operating orders of magnitude faster, and at far lower power, than traditional digital circuitry \cite{chua71,strukov08,review1,review2,ProbComp5}. Since memristive behavior is associated to a 1-port device which is current- or voltage-controlled, these components support volumetric memory scaling for dense storage capacity, with the possibility of being simply operated on a crossbar array  \cite{chung,yang,zidan,xia,li,li2,li3,hu,prezioso,jiang,sawa2,wagenaar,ruitenbeek,pickett,mahne,kim}.
Moreover, recent results suggest that networks of memristive devices \cite{Caravelli2015,Caravelli2016rl,Caravelli20192} can be used for various tasks in analog computing, and exhibit chaos and system switching phenomena \cite{caravelliscience,zdenka}.

One advantage of these nanoscale devices is that with a displacement of a few atoms, we have possibly access to resistive states at ultimately low energy costs. However, device engineering lacks
the complete quantum theoretical characterization of resistive switching when this is driven by a conductance change due to a filament formation, also considering that the nature of the filament formation is inherently stochastic. Although some analytical methods exist, these are restricted to the semiclassical domain \cite{carmilano}.

Filament formation, e.g. the nonequilibrium process of chemically forming a filament in a narrow region \cite{aono,cheng,agrait,makk,scheer}, is a mechanism for resistive devices. However, with the increasing capability of scaling down to a few atoms, and the ability to measure quantum effects in these filaments at a few Kelvins, new techniques are necessary to capture the phenomenology of these devices. 

Quantum transport, on the other hand,  has been extremely successful in the study of nanoscale devices \cite{cuevas10,Datta95,DiVentra}, whether using the Kubo \cite{Kubo57} or Landauer approach \cite{Landauer57,Fisher81,Economou81,Imry86,Buttiker85,Buttiker86,Buttiker88a,Buttiker88b,Buttiker88c}. 
For the case of non-filamentary switching, nanoimaging techniques have shown recently that charge density waves emerge when two carriers of opposite charge are allowed in the narrow region, for instance in the case of perovskite nickelates doped with Hydrogen \cite{abate}. The present paper focuses in particular on nanoscale filamentary switching.
Recently, state-of-the-art experimental results
managed, via
Andreev reflection processes \cite{Andreev64a,Andreev64b,Andreev65,Blonder82,ludoph}, to measure quantum effects in niobium pentoxide \cite{halbritterj,molnar,liuc,wylezich1,wylezich2,marchenkov}, emerging at the filament terminals when superconducting electrodes are utilized \cite{moussy,vinet,shen,leseuer,scheerx,belzig,schulz}. 
In particular, in certain cases, the quantum PIN code, i.e. the transmission probabilities of each conduction channel contributing to the conductance of the nanojunctions can be obtained \cite{halbritter} using Andreev reflections. 
These studies suggest that atomic-size metallic filaments are indeed the source of high conductance in niobium pentoxide (the so-called ON state).
These recent studies motivate the theoretical investigation of transport in heterogeneous filamentary structures \cite{aono,cuevas,gubicza1,gubicza2}. 

However, immediately one faces an obstruction.  Previous studies focusing on filament formation, directly simulating filament formation, have used either continuum models and/or well-defined lattices,  or classical Kirchhoff and Ohm's laws to calculate the resistance. On the other hand, quantum mesoscopic transport techniques are based, typically, on the assumption of continuity, e.g. the conductor acts as a waveguide for the charge carriers \cite{cuevas10,Datta95,DiVentra}. As such, the study of quantum transport on this mesoscopic network is a challenging task already at the modeling level, even before any calculation can be attempted.
\begin{figure}[t!]
    \centering
    \includegraphics[scale=1.1]{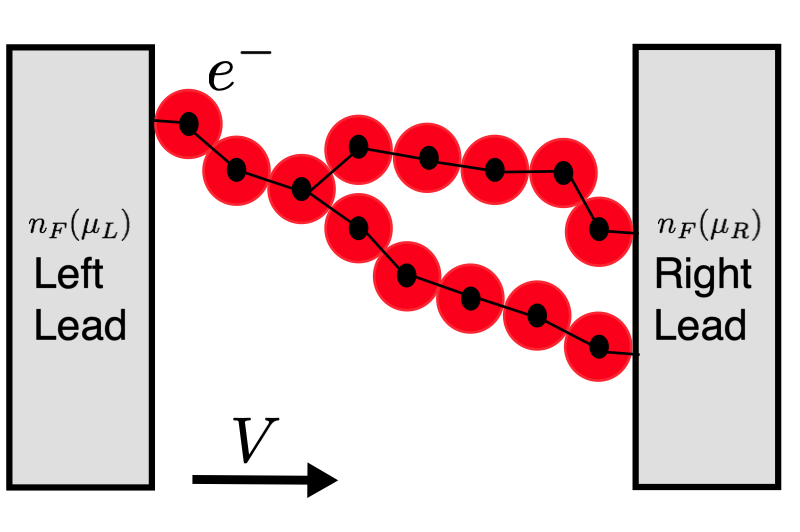}
    \caption{Sketch of filamentary switching for a few atoms. The structure is far from continuous, and the charge carriers hop across the structure as induced by the voltage. The formalism we use is the one by Landauer-B\"{u}ttiker, with two reservoirs (leads).}
    \label{fig:structure}
\end{figure}

For this reason, the general approach we consider in this paper is represented in Fig. \ref{fig:structure}. We consider a left and right lead and the dynamical formation of a filament in the region. We assume that the filament formation is classical, and on a timescale much longer than the effective speed of the electrons across the filament. The filament is composed of an assembly of $N$ atoms, with overlapping bands. 
However, the scattering process of the electron across the filament is quantum mechanical, and thus one has to consider the scattering amplitude {$T_{ij}$} from the lead $i$ to the lead $j$ quantum mechanically.

However, over the past three decades, the study of electrical conductors has evolved considerably from a macroscopic level of description to the nanoscale. 
Assuming that the characteristic length of the quantum correlations is more or less of a few atoms (e.g. dephasing or Fermi lengths are small), we can assume coherent quantum transport of electrons. The setup we study is the typical one of mesoscopic transport \cite{Kubo57}, e.g. our system has two leads, denoted as left and right, and a voltage applied to them.
Such a picture has been extremely successful in the study of molecular and atomic point contacts \cite{cron, schirm,Olesen94,Krans95,Scheer97,Rodrigues00}, carbon nanotubes \cite{Frank98,Martel98}  or graphene \cite{Titov06,Cuevas06,Gonzalez08,Peres06,Tombros11}. More recently,
atomic point contacts have been also investigated for resistive switching \cite{gong,torrezan} and there has been an interest in generic quantum wires \cite{quantumwires} or Josephson junction networks  \cite{trombettoni}.

The idea that the discreteness and yet heterogeneity of the structure might play a role has been investigated as early as the 30's by Pauling \cite{pauling,ruedenberg,montroll1,montroll2}.
Thankfully, the study of scattering on discrete structures has obtained some considerable attention over the past decades (the so-called \textit{quantum graphs}) \cite{roth,kottos,kottos2,schanz,schmidt,andrade,berkolaiko,berkolaiko2,bolte,kostrykin,kostrykin2,chadan,kostrykin3,economou}, driven by the study of quantum chaotic systems. Mathematically, the overlap between the atoms can be described as a graph \cite{diestel,kuchment}, with the atoms represented as vertices and the overlap as edges of such graph.

The assumption is that electrons on the left or right leads are asymptotically free and incoming and outgoing states \cite{Landauer87,Kawabata89,Landauer91,Landauer92,Buttiker92}. The electrons can propagate freely or in a constant potential $u_{ij}$ along the edge of the graph (a one dimensional structure), whose distance is assumed to be constant {$\ell$}. Then, the problem of transport can be described as the problem of calculating the scattering matrix which transforms the incoming into outgoing free states on these ``quantum'' graphs \cite{waltner,ragoucy,caudrelier,gnutzmann,andrade1,blumel1,blumel2,luz,andrade2,andrade3,andrade4}.
Intuitively, Kirchhoff's currents conservation for all the nodes of a circuit, and which is classically written in terms of the directed incidence matrix of a graph as $B^t =\vec i=0$ \cite{diestel,Caravelli2016rl}, has to be replaced by a ``quantum'' Kirchhoff law in which a unitary matrix has to be inserted in the sum over currents at a node.  In quantum graphs, these unitary operators can be obtained by the scattering process a local one-body Hamiltonian.

For an entire network connecting two leads, such a unitary matrix contains the information about the transmission {amplitudes};
at this point, we can use the transmission {amplitudes} within the context of a Landauer approach. In fact, the probably best-known result in the theory of quantum transport obtained using the scattering matrix approach is the Landauer formula~\cite{Landauer57}, who was the first to use scattering matrices to describe transport problems. For multi-leads, we use a generalization of this approach, also known as the Landauer-B\"uttiker formula. Specifically, we will treat our system as a particular case of a multi-lead system, where however some of the leads are identical. In the case of several channels, the expression for the conductance
\begin{equation}
	G =\frac{2e^2}{h}\sum_{n,m} T_{nm}
	\label{eq:GN3}
\end{equation}
contains the sum of transmission probabilities $T_{nm}$ from one mode (channel) to another. The expression above will have to be adapted to the case of quantum graphs which we discuss below. In the more specific case of a filament, we will see that we can have $K$ channels attached to the left and $K^\prime$ channels attached to the right lead.

The approach we use here is different from the case of the non-equilibrium Green's function \cite{Datta95,DiVentra} in various ways. First, the scattering process is on a discrete structure rather than a continuous one, as it is commonly assumed in analytical calculations \cite{Buttiker90b,Beenakker91,Lesovik89b,Buttiker90}. Moreover, we make the simplifying assumption that the electrons are faster than the typical timescale of the filament formation. As such, we consider the equilibrium Green's functions on a particular structure. This is motivated by the fact that the typical formation of a filament in these materials is of the order of a few milliseconds, in which this annealed approximation makes sense.

The paper is organized as follows. We first review the standard Landauer-B\"{u}ttiker approach for the multichannel scattering. We then introduce the setup of this paper and the construction of the scattering matrix based on quantum graphs. We then discuss the molecular dynamics simulations that we employed for the stochastic filament formation, and give the assumptions on the dielectric. Conclusions follow.

\section{Standard Landauer approach and graph scattering matrix}
The approach we consider here is the one introduced in \cite{Buttiker85}.

\subsection{Landauer-B\"{u}ttiker formalism}
To describe a quasi-one-dimensional coherent conductor, we first consider a purely one-dimensional problem \cite{Datta95,DiVentra}.
We consider for instance electrons with energies up to a certain scale $\mu$ in a scattering state, moving (for instance) left to right,
in a Lippmann-Schwinger scattering states \cite{Lippmann50,Landau04BookV3} of the form

\begin{equation}
	\Psi_{ L,  E}(x) =
	\begin{cases}
		e^{i k x} + r(E) e^{-i k x},	& x \to -\infty, \\
		t(E) e^{i k x},			& x \to \infty,
	\end{cases}
	\label{eq:scatterL}
\end{equation}
where we introduced  $k = \sqrt{2m E}/\hbar$, and we assume the state above to be normalized. The quantities $r(E)$ and $t(E)$ are the reflection and transmission {amplitudes} respectively. For right-moving states, we have an analogous state to the one of eqn. (\ref{eq:scatterL}). The density of states in one dimension is given by the expression	$\nu(E) = \frac{dk}{dE} = \frac{m}{\hbar^2 k}$.
Because of the electron statistics, the probability of having an electron at a certain energy $E$, is given by
\begin{equation}
    f(E)=\frac{1}{e^{(E-\mu)/\kappa T}+1}.
\end{equation}
where $\mu$ is a chemical potential, which in the Landauer approach is assumed to be dependent on the voltage potential at the lead.
If we have two leads, then, the left and right leads depend on two chemical potentials $\mu_L$ and $\mu_R$, which we assume to be dependent on the voltage at the two electrodes, anodes and cathodes.
The total current is the sum of the left and right contribution, given by the expression \cite{Landauer87,Kawabata89,Landauer91,Landauer92,Buttiker92}
\begin{eqnarray}
	I(V) &=& I_{ L}(V) + I_{R}(V)  \nonumber \\
	&=&
	\frac{2e}{h} \int_{-\infty}^\infty dE (f_L(E)-f_R(E)) T(E),
	\label{eq:totalCurrentB}
\end{eqnarray}
where $\frac{2e}{h} =G_0/e$, where $G_0$ is the conductance quantum.
The transmission probability from left to right, $T=|t|^2$, is equal to the transmission probability from right to left, $T=T'=|t'|^2$ because of unitarity, and then in this sense, all we need to know is the scattering amplitudes for the system given the left and right scattering. 

The quantity $T(E)$ is the sum over all channels, $T(E)=\text{Tr}(\tau^\dagger(E) \tau(E))$, where $\tau(E)$ is the transmission matrix of the system.

We see that within this approach, we only need to solve the scattering process across the structure, whether this is a continuum or discrete. 
As we have said above, in this paper we consider only local effective one-dimensional movements of the electrons in between the underlying atoms forming the filaments. For this purpose, then, we need to introduce the graph scattering process associated with our generalized Lippman-Schwinger states.

In general, if the system has many leads, the total transition probability can be written as 
\begin{eqnarray}
	I_k(V) &=& 
	\frac{G_0}{e} \int_{-\infty}^\infty dE (f_L(E)-f_R(E)) \sum_j T_{kj}(E)
	\label{eq:totalCurrentC}.
\end{eqnarray}
where $T_{kj}$ is the transition probability from the lead $j$ to the lead $k$. 

If the leads $k$ are then attached to the same contact, we can write 
\begin{eqnarray}
	I_{tot}(V) =
	\frac{G_0}{e} \int_{-\infty}^\infty dE (f_L(E)-f_R(E))  \sum_{jk}T_{kj}(E)
	\label{eq:totalCurrentD}.
\end{eqnarray}
Since in quantum graphs we do obtain the quantities $T_{kj}$, we will use this expression. At small temperatures and for $\mu_L-\mu_R\approx e V$, this reduces to
\begin{eqnarray}
	I_{tot}(V) =
	\frac{G_0}{e} e  \sum_{jk}T_{kj}(E_f) V= G_0(\sum_{jk}T_{kj}(E_f)) V,
	\label{eq:totalCurrentE}
\end{eqnarray}
where $E_f$ is the Fermi energy for the system and $V$  the applied voltage $G_0\approx $ (12900 Ohms)$^{-1}$, the quantum of conductance. This equation cannot be directly applied in the theory of quantum graphs in a sense we will see in the next subsection, but a simple modification will be sufficient to allow us to use it.

\subsection{Adapting the LB formula to quantum graphs}

Before we discuss the theory of quantum graphs, it is worth discussing what one can calculate and how the results should be interpreted. Quantum graphs are a generalization of the one-dimensional Schr\"{o}dinger equation, in which single particles enter along a one-dimensional path between the scattering nodes, and the electron can move across a heterostructure.  In this sense, particles do not have any transverse modes, and the scattering process is purely one-dimensional.  The starting point is a certain graph representing the atomic structure and the overlapping orbitals. In such a graph, the nodes represent where the particles scatter and the edges are the one-dimensional paths in between scattering events.

What one can calculate with this mathematical construct is the transmission probability $T_{ij}(E)$, e.g. the chances that the particle being coherent along the mesoscopic region enters the region through a certain node $i$ and leaves from node $j$.
Now let us assume that our scattering region, approximated as a graph, is attached via $N$ nodes on the left lead and $M$ nodes on the right lead.

If the particle incomes from the left lead, and want to calculate the right-moving current $I^L$, then we assume that will have \text{classical} probability $p^L_i(E)$ to enter at node $i$, satisfying $\sum_{i=1^N}p^L_i(E)=1$. Then, the total current, following a similar argument to the one of Landauer, is given by
\begin{eqnarray}
I^L(V)=G_0/e \int dE f_L(E,V) \sum_{i=1}^N \sum_{j=1}^M T_{ij}(E) p_i^L(E).
\end{eqnarray}

\begin{figure}
    \centering
    \includegraphics[width=0.45\textwidth]{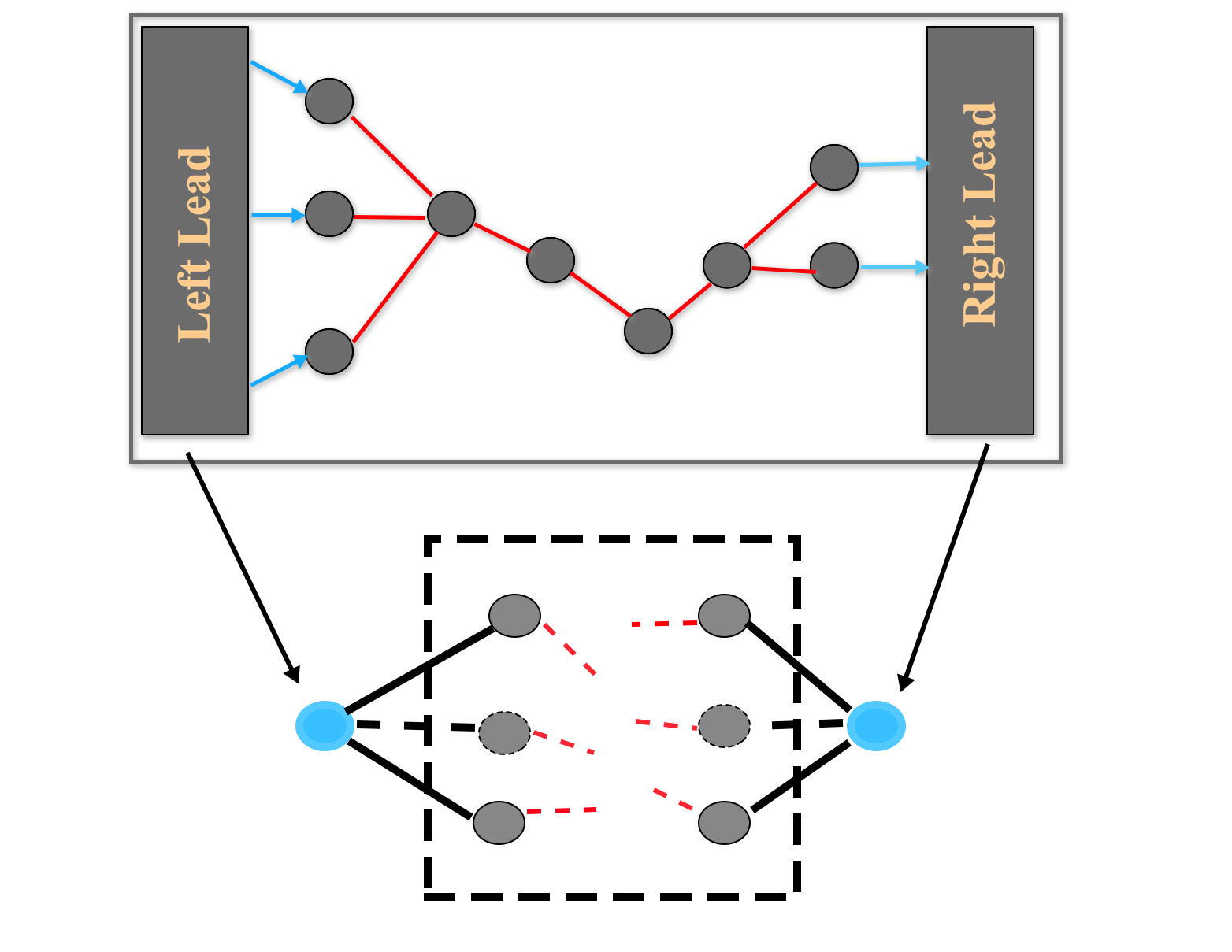}
    \caption{Reduction of the lead connections to a single channel and incoming-outgoing Lippman-Schwinger asymptotic states. }
    \label{fig:lippsch}
\end{figure}

A similar argument can be made for the left moving particles incoming from the right lead, obtaining
\begin{eqnarray}
I^R(V)=-G_0/e \int dE f_R(E,V) \sum_{i=1}^N \sum_{j=1}^M T_{ij}(E) p_j^R(E),
\end{eqnarray}
from which we get the total current expression
\begin{eqnarray}
I(V)&=&I_L(V)+I_R(V)\\
&=&G_0/e \int dE \sum_{i=1}^N \sum_{j=1}^M\Big(f_L p_i^L-  f_Rp_j^R\Big)T_{ij}\nonumber
\end{eqnarray}
where we have suppressed the functional dependence for simplicity. 
We note that this expression is similar to
the one of the Landauer-B\"uttiker formula, except for the correcting factors $p_i^L$ and $p_j^R$. The method above is, however, time-consuming, as it requires the evaluation of $ML$ channels' probabilities. 
We use instead the following trick, commonly used in the literature \cite{dragons1,dragons2,dragons3} of quantum dragons to study the transmissions to a single channel. Instead of repeating the process $ML$ times, we attach two extra nodes to the graph attached to the boundary nodes (connected to the leads). Using this method, we reduce our problem to a single Lippman-Schwinger asymptotic in and out states. This is shown in Fig. \ref{fig:lippsch}.

\subsection{Quantum graphs: quantum scattering on graphs}

In the Landauer-B\"{u}ttiker approach all one needs to know is the scattering amplitudes on the underlying structure. As mentioned earlier, we take the point of view in which the scattering occurs along one-dimensional paths connecting the underlying atoms. This is a departure from previous works in quantum transport, but it can be considered as a model approximation to calculate the scattering amplitudes as a function of the incoming energy.
It is thus worth formalizing the process, as it is not a standard approach in transport. The goal of this and the next few sections is to introduce the formalism that we use.

A graph $\mathcal G(V,E)$ is the tuple given by the set of vertices
vertices $V(\mathcal G)=\{1,\ldots,n\}$  and the set of edges between nodes,
$E(\mathcal G)=\{e_{1},\ldots,e_{l}\}$, where each edge is a pair of vertices
\cite{diestel}.

\begin{figure}
  \centering
 
  \includegraphics*[width=0.6\columnwidth]{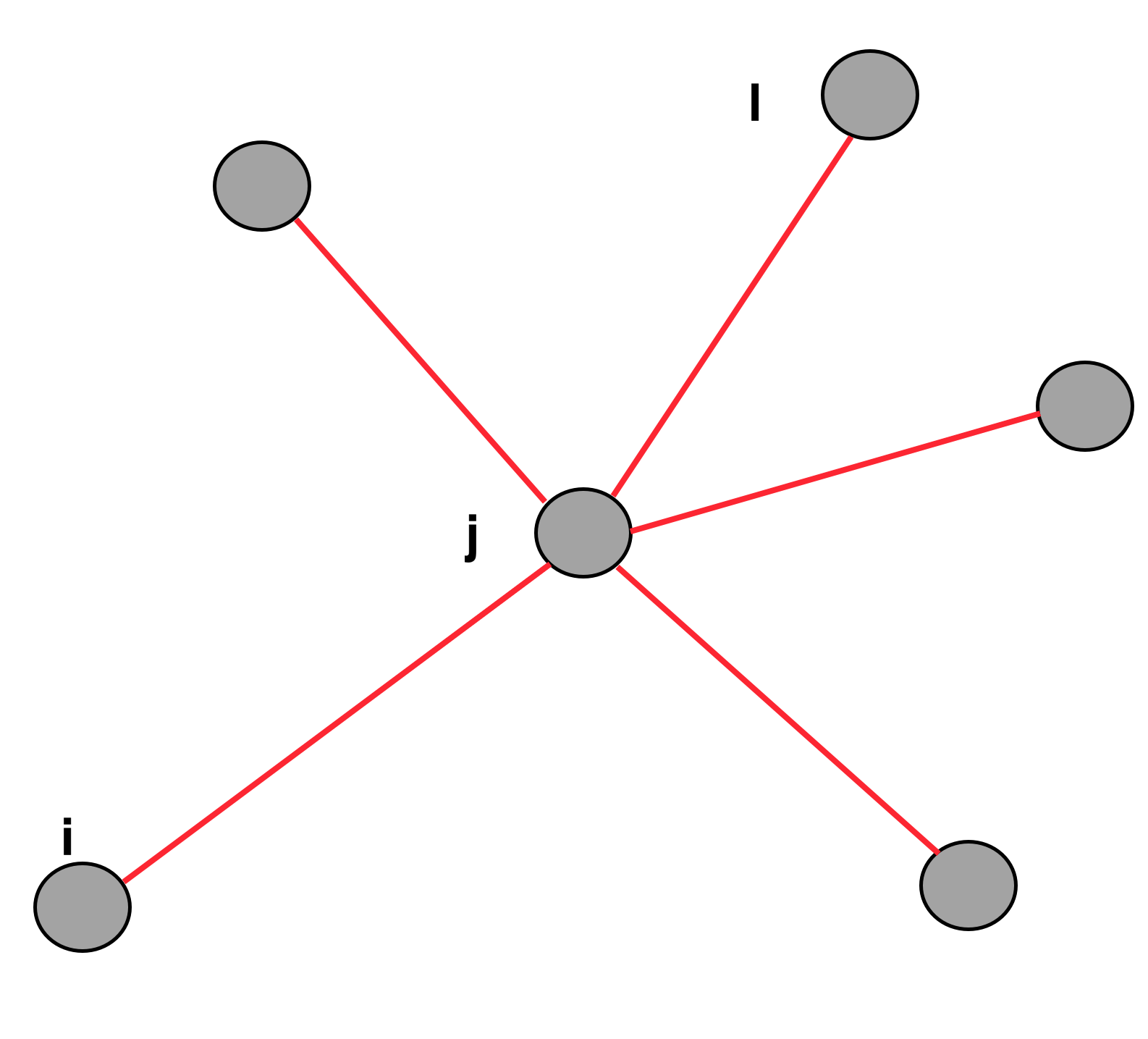}
  \caption{
    Locally, any graph looks like a star graph.}
    \label{fig:fig1}
\end{figure}

In the near vicinity of any node, a quantum graph always looks like a star graph of the type shown in Fig. \ref{fig:fig1}.

The graph connectivity is encoded in the  $n \times n$ adjacency matrix $A(\mathcal G)$, $A_{ij}$ is one if the vertices $i$ and $j$ are connected and zero otherwise. The set $E_i=\left\{j:\{i,j\}\in E(\mathcal G)\right\}$ is the neighborhood of the vertex $i\in V(\mathcal G)$ (e.g. at distance one).
We denote with $E_{i}^{k}= E_{i}\setminus \e{k}$ the set of neighbors
of the vertex $i$ except for $k$. The degree of the vertex $i$ is defined as $d_i=|E_i|=\sum_{j=1}^{n} A_{ij}(\mathcal G)$. These definitions are standard in graph theory and are called \textit{simple} or \textit{discrete} graphs. 

However, since our graph is embedded in space, it is useful to also consider \textit{metric} graphs, in which the notion of distance is encoded.
To discuss quantum graphs, it is necessary to equip the graphs with a
metric.
A \textit{metric graph} $\Gamma(V,E)$ is a graph in which every edge is assigned a certain length $\ell_{e_{s}}\in(0,+\infty)$, defining $\boldsymbol{\ell}=\{\ell_{e_{1}},\ldots,\ell_{e_{l}}\}$.

Given this definition, we call leads those edges which have a semi-infinite length $e_{s}$ ($\ell_{e_{s}} = + \infty$); these edges are associated with the Lippman-Schwinger states (in the literature, these are also called ``open {quantum} graphs").

\subsubsection{Basic theoretical framework}
Quantum graphs are metric graphs in which the Schr\"odinger operator is defined along the edges of the graph. At every vertex, we have boundary conditions (BCs). The triple
$\{\Gamma(V,E),\bold{H},\bold{bc}\}$ with $H$ a differential operator and
$\bold{bc}$ a set of BCs defines a quantum graph.
For instance, along a certain edge $\{i,j\}$, the free Schr\"{o}dinger operator $H_{ij}=-(\hbar^2/2m)d^2/d{x_{e_{ij}}}^2$ is associated to the eigenvalue equation on the edge $\e{i,j}$ is given by
\begin{equation}
  \label{eq:eigen}
  - \psi_{\e{i,j}}''(x) = k^2 \psi_{\e{i,j}}(x),
\end{equation}
where $k=\sqrt{2mE/\hbar^2}$, $m$ is the mass of the particle, $E$ is its energy, and
$\psi_{\e{i,j}}$ is the wave function. We call \textit{naked quantum graphs} the triple in which $H$ is the free evolution, e.g. there is no potential, and \textit{dressed quantum graphs} those in which edge potentials are present. 

In general, formally, the total wave function of the system is a collection of local wavefunctions assigned to the edges. We have
\begin{eqnarray}
    \Psi=\begin{pmatrix}
    \psi_{e_1}(x_{e_{a_1 b_1}})\\
    \vdots\\
        \psi_{e_m}(x_{e_{a_m b_m}})\\
    \end{pmatrix}
\end{eqnarray}
where the pairs $(a_m b_m)$ represent an edge, e.g. a pair of nodes, and $m$  is the total number of edges. In general, $x_{e_{ij}}$ is a local parameter such that for $x=0$ we have the wave function on the vertex $i$, while $x=\ell$ we have the wavefunction on the vertex $j$. We will often write $e_k$ to denote a certain edge, dropping the two vertices notation unless necessary. The Hamiltonian operator 
\begin{eqnarray}
H_{e_k}(x_{e_j})=-\frac{\hbar^2}{2m} \frac{d^2}{dx_{e_j}^2}+V(x_{e_j})
\end{eqnarray}
is the standard 1-d particle Hamiltonian on the compact domain $[0,\ell_j]$.

In the theory of quantum graphs, vertices are interpreted as zero-range potentials, where boundary conditions need to be supplied. The collection of boundary conditions is, for every node, the continuity and differentiability of the wave functions. If $\{e_1,\cdots,e_k\}$ are the edges adjacent to a certain vertex $n$, then we must have
\begin{eqnarray}
(1)& &\ \ \ \ \  \psi_{e_1}|_{n}=\psi_{e_2}|_{n}=\cdots=\psi_{e_k}|_{n}  \label{eq:k1}\\
(2)& &\ \ \ \ \  \sum_{j=1}^k \partial_{x_{e_j}} \psi_{e_j}|_{n}=0 \label{eq:k2}\\
    & &\text{(either at $x_{e_j}=0$ or $ x_{e_j}=\ell_j\}$ depending on the definition} \nonumber 
\end{eqnarray}
In the case of zero potential for the edge, the wavefunction is generically of the form
\begin{eqnarray}
    &-&\frac{d^2 \psi_{e_k}}{d x_{e_k}^2}=k^2 \psi_{e_k}\nonumber \\
    & &\ \ \ \rightarrow \psi_{e_j}(x_{e_j})=c^+_{j} e^{i kx_{e_j}}+c^-_j e^{-i kx_{e_k}} \label{eq:solk}
\end{eqnarray}
A Lippman-Schwinger state is a particular choice of an edge as in the incoming 
``left" state, and accordingly a right state \cite{Lippmann50}. For instance, a ``left" (incoming) state on the edge $i$ and a ``right" (outgoing) state on the edge $f$ is given by the pair
\begin{equation}
	\psi_{ LS,  \{e_i,e_f\}} =
	\begin{cases}
		e^{i k x_i} + r(E) e^{-i k x_i},	& x_i \to -\infty,\\
		t(E) e^{i k x_f},			& x_f \to \infty.
	\end{cases}
	\label{eq:scatterLg}
\end{equation}
We write now the left and right in quotation simply because our system is not simply one dimensional any more, but the notation stands.

\begin{figure}
  \centering
  \includegraphics[width=0.9\columnwidth]{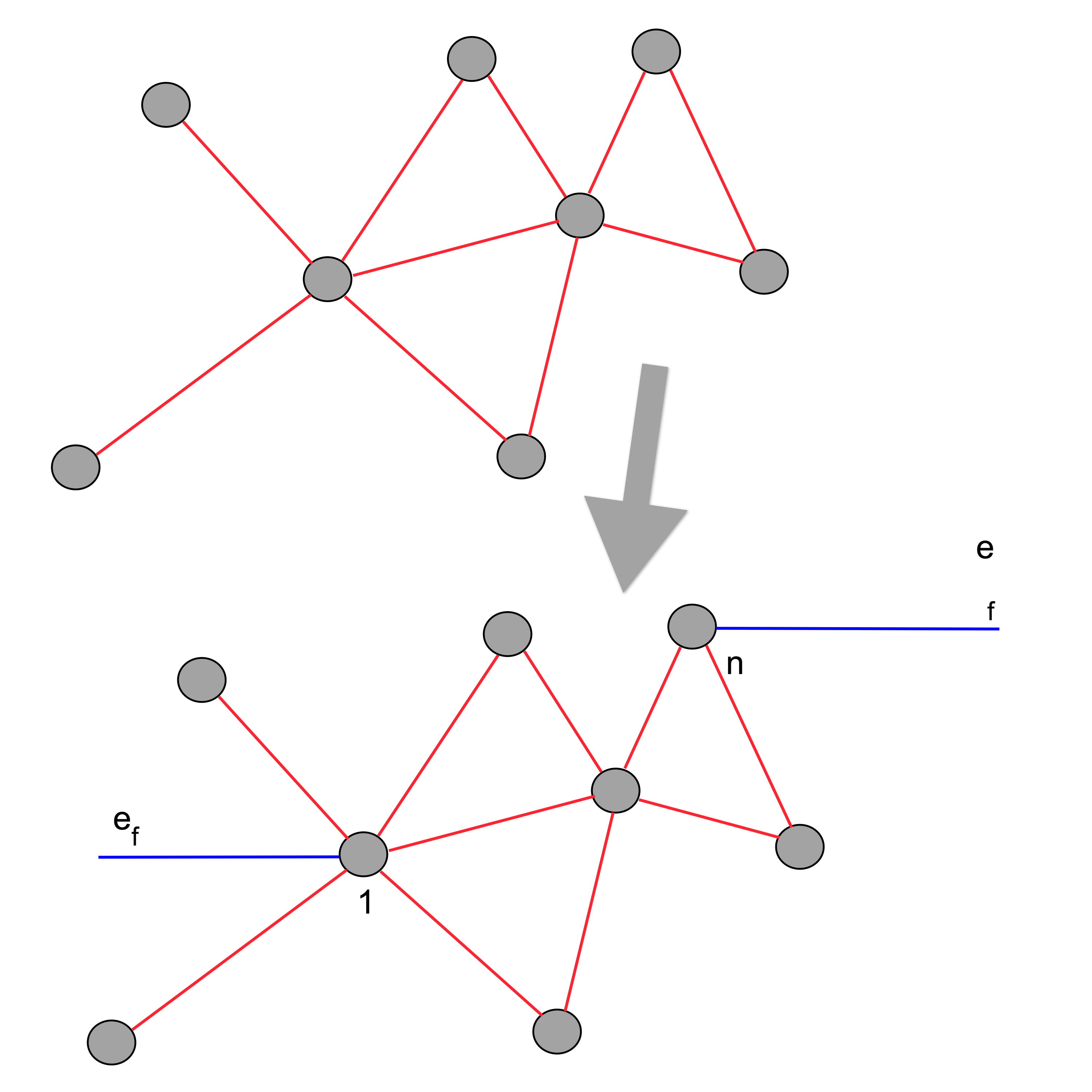}
  \caption{Graph with two leads added turning it into an open quantum graph.}
  \label{fig:fig2}
\end{figure}

Eqs give a solution of the quantum graph equations. (\ref{eq:k1})-(\ref{eq:solk}) plus the external boundary conditions given by the Lippman-Schwinger asymptotic states of Eq. (\ref{eq:scatterLg}); these are called ``open'' quantum graphs, and are represented in Fig. \ref{fig:fig2}. In this state, the quantum graph formalism is equivalent to the scattering approach necessary for the Landauer-B\"{u}ttiker approach, or Eq. (\ref{eq:totalCurrentB}).
While Green's function approach is unitarily equivalent to the Schrödinger approach for quantum graphs, the former can be automatized by using the adjacency matrix, which is the technique we use here to study transport numerically and we will review below.


\subsubsection{Green's function approach}
An important quantity in the theory of quantum scattering in general, and in particular in the theory of quantum graphs, is the Green's function approach (GFA).
The Green's function for a certain edge is defined as 
\begin{eqnarray}
    [E-H(x_f)]G(x_f,x_i; E)=\delta(x_f-x_i),
\end{eqnarray}
which is the Green's function of a 1-dimensional quantum particle. Formally, the solution of the equation above can be written exactly, and we invite you to read the review \cite{andrade} on the subject. We only wish to add that there are a few methods to calculate Green's function exactly, including the exact solution of the differential equation, using a spectral representation of the exact solution, or alternatively using a Feynmann path integral representation.

The approach that we emphasize here is the path integral approach. It is in fact remarkable that in fact for quantum graphs, the semiclassical approach introduced by Gutzwiller, or also Gutzwiller trace formula, is exact \cite{andrade}. As such, Green's function can be written \textit{exactly} as the sum over all classical periodic orbits.
Specifically, the Green's function can be written as 
\cite{kostrykin2,kostrykin} (see also \cite{andrade})
\begin{eqnarray}
    G(x_f,x_i; E)=\frac{m}{i \hbar^2 k} \sum_{\rm sp} W_{\rm sp} e^{\frac{i}{\hbar}  S_{sp}(x_f,x_i; k)},
\end{eqnarray}
where $\rm sp$ stands for scattering paths: these are \textit{all} the paths that begin in $x_i$ and end in $x_f$. Clearly, the sum above is possibly over an infinite number of paths, given by the fact that a particle can be in principle reflected in a countable many ways on an edge before leaving the graph. This sum can be performed exactly. The quantity $W_{\rm sp}$ is the quantum amplitude of the path: for each time a particle is reflected, it is multiplied by a factor $r_j(E)$ on the specific edge if it is reflected, and if transmitted it is multiplied by $t_j(E)$. The quantity $S_{\rm sp}$ is instead the classical action of the particle over that specific path: for each edge $j$, the particle has traversed along the path, we need to add $S_j=\ell_j k$. Let us make a quick example which shows why this is useful.

\begin{figure}
    \centering
    \includegraphics[scale=0.5]{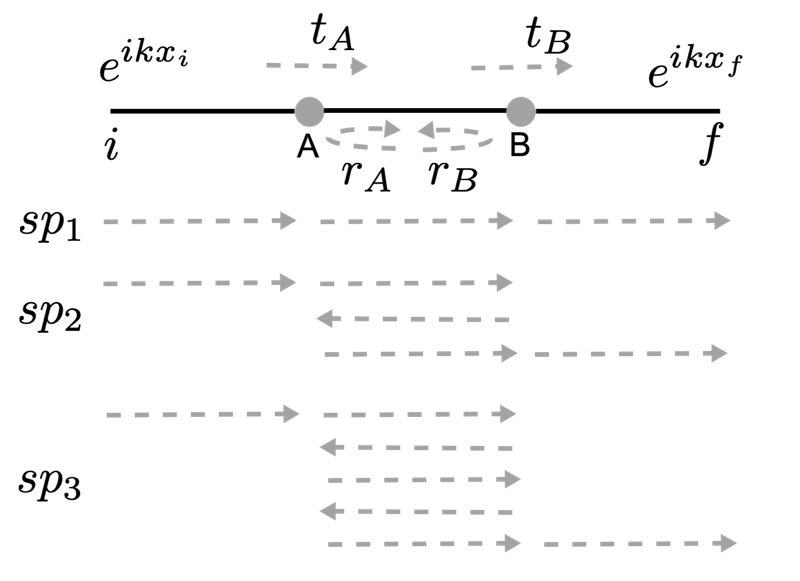}\\
        \includegraphics[scale=0.5]{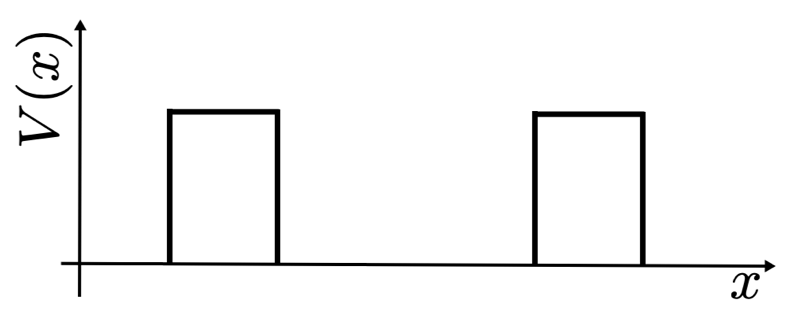}
    \caption{\textit{Top:} The simplest graph which has a nontrivial transmission probability. The quantities $t_A$ and $t_B$ are the transmission {amplitudes} for the incoming wave function, while $r_A$ and $r_B$ are the reflection {amplitudes}. The two ``open" edges, on the left and right, are the semi-infinite leads associated with the asymptotic Lippman-Schwinger states. Calculating the Green's function requires the enumeration of all possible paths connecting the two asymptotic states. \textit{Bottom:} The equivalent shape of the potential associated with the top graph. }
    \label{fig:scattering}
\end{figure}

Let us use now the following notation: for a vertex $j$, $r_j$, and $t_j$ are the reflection and transmission probabilities associated with the wavefunction parametrization. As an example, let us look at Fig. \ref{fig:scattering}. We have two incoming and outgoing Lippman-Schwinger {states}, and one internal edge bordered by the vertices $A$ and $B$, whose distance we assume to be simply $l$. It is not hard to see that any scattering path can be written in the forms of $sp_1,\cdots ,sp_3,\cdots$: if a particle is reflected at $A$, then because the edge $i$ is infinite it will never return back, and it cannot be part of a scattering path.

The amplitude $W_{1}$ associated with $sp_1$ of Fig. \ref{fig:scattering} (top) is simply $t_A t_B$, while the classical action is given by $S_{sp1}=k l+kx_i + k x_f$. For $sp_2$, we have $W_2=t_A r_B r_A t_B$, and $S_{sp1}=3 k l+kx_i + k x_f$. For $sp_3$, we have $W_2=t_A  (r_B r_A)^2 t_B$, and $S_{sp1}=5 k l+kx_i + k x_f$. Any path can be written in this form, and it is not hard to see that the Green's function can be written as a geometrical series. We obtain then
\begin{eqnarray}
    G_{fi}(x_f,x_i; k)=\frac{m}{i \hbar^2k} \frac{t_A t_B e^{ikl}}{1- r_A r_B e^{2i k l}} e^{ik(x_f+x_i)}.
\end{eqnarray}
What is interesting about this expression is that we can read out the transmission amplitude for the whole system, $T_{fi}$, given by
\begin{eqnarray}
    T_{fi}=\frac{t_A t_B e^{ikl}}{1- r_A r_B e^{2i k l}}.\label{eq:transmfi}
\end{eqnarray}
At this point, we see that this total scattering amplitude has the exact same form as the
as the scattering of a quantum particle in one dimension in a potential $V(x)$ of the form of Fig. \ref{fig:scattering} (bottom). This example shows that the quantum graph technique is a legitimate option to calculate the scattering amplitude in a mesoscale system under the single-body approximation, although it also provides a generalization to more complex structures. 
In general, the Green's function can then be written as
\begin{eqnarray}
    G_{fi}(x_f,x_i; k)=\frac{m}{i \hbar^2k} T_{\Gamma} e^{ik(x_f+x_i)}.
    \label{eq:gfa}
\end{eqnarray}
where $T_{\Gamma}$ is the transmission, which is what we aim to calculate.

In what follows, we will also use another technique to evaluate the total transmission probability which has been obtained recently and is based on the adjacency matrix of the graph.

\subsubsection{Adjacency matrix approach}

One of the main ingredients of the GFA is the local origin of the total scattering amplitude and the fact that the scattering at each vertex can be defined locally.

For each vertex, we generalize the local reflection and transmission amplitudes into a tensor, $\boldsymbol{\sigma}_{j}$, depending on each vertex $j$ label.
Locally, the scattering amplitudes are in fact those of a star graph, a property which has been very fruitful in the study of quantum graphs. More details on the connection between the boundary conditions at each star, quantum probability flux conservation, and self-adjointness can be found in \cite{kostrykin2}. We will now take a more practical approach, introduced in \cite{andrade4}.
For a star graph, let the local wavefunction be given by 
$\Psi(j)=\left(\psi_{\{j,1\}}(j),\ldots,\psi_{\{j,n\}}(j)\right)^{T}$.
The most general BCs that are consistent with the self-adjoint condition
\cite{andrade} are totally defined by two $d_{j} \times d_{j}$ matrices.
At each node, in fact, given two matrices $\mathcal{A}_{j}$ and $\mathcal{B}_{j}$, the boundary conditions we defined earlier can be  generalized to
\begin{equation}
  \label{eq:bc}
  \mathcal{A}_{j} \Psi(j)+ \mathcal{B}_{j} \Psi'(j) = 0.
\end{equation}
Note that $\mathcal A_{j}$ is not the adjacency matrix of the graph, but represents the nodal boundary condition.

The only requirement is that the matrix
$\mathcal{A}\mathcal{B}^{*}$ is self-adjoint, while the
matrix $\left(\mathcal{A}_j,\mathcal{B}_j\right)$
has the maximal rank $d_{j}$ \cite{andrade4}. We now want to see that defining a local scattering matrix is equivalent to the boundary conditions.
The boundary conditions of eqns. \eqref{eq:bc} can be
in fact determined by considering a plane wave on the edge $e_{i,j}$ incoming into
the vertex $j$, which we assume to be of degree $d_{j}$. 

\begin{figure}
    \centering
    \includegraphics[scale=0.5]{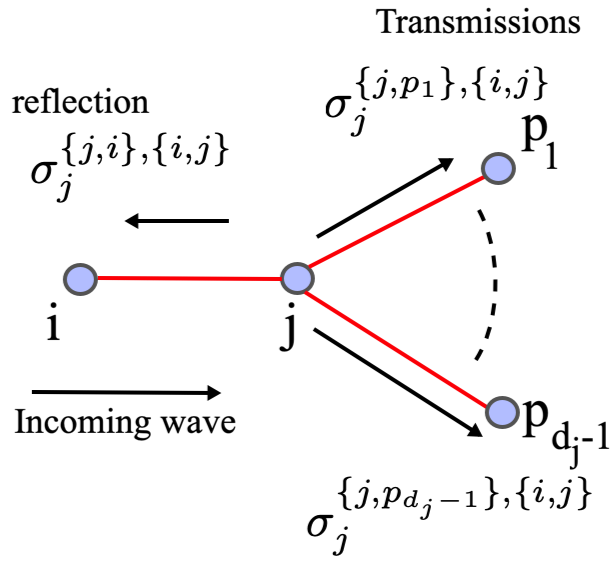}
    \caption{Transmission and reflection probabilities for an incoming wave from $i\rightarrow j$.}
    \label{fig:transfref}
\end{figure}

Let us introduce the matrix $\boldsymbol{\sigma}_j$, which is related to the reflection and scattering amplitudes as $\sigma_{j}^{[\e{j,i},\e{i,j}]}(k)=r_{j}^{[\e{j,i},\e{i,j}]}(k)$
and
$\sigma_{j}^{[\e{j,p},\e{i,j}]}(k)=t_{j}^{[\e{j,p},\e{i,j}]}(k)$,
where $r_{j}^{[\e{j,i},\e{i,j}]}$ is the reflection in vertex $j$ for the wave incoming from $i$, while $t_{j}^{[\e{j,p},\e{i,j}]}(k)$ the transmission through vertex $j$, for the wave incoming from $i$ and scattering to $p$. Unitarity typically requires that these are not independent quantities.

Given the definition above, the ``local" Lippman-Schwinger states at the node $j$ are given by
\begin{align}
  \label{eq:wavef}
  \psi_{\e{i,j}}(x) = {}
  & e^{- i k x}
    + \sigma_{j}^{[\e{j,i},\e{i,j}]}(k) e^{i k x},\nonumber \\
  \psi_{\e{j,l}}(x) = {}
  &  \sigma_{j}^{[\e{j,l},\e{i,j}]}(k) e^{i k x}.
\end{align}
A direct application of eqn. \eqref{eq:bc} gives
\begin{equation}
  \label{eq:sigma}
  \boldsymbol{\sigma}_{j} (k) =  -
  (\mathcal{A}_{j} + i k \mathcal{B}_{j})^{-1}
  (\mathcal{A}_{j} - i k \mathcal{B}_{j}).
\end{equation}
A few comments are in order. First, one can either set the matrix $\boldsymbol{\sigma}_{j}$ or the boundary conditions \cite{kostrykin2}. Second, the scattering matrix depends 
on $k$, and thus on the energy $E_k=\frac{\hbar^2 k^2}{2m}$, in a non-trivial manner (although certain boundary conditions are $k$-independent \cite{bolte}).  

The conservation of probability is encoded in the unitarity of the matrix $\boldsymbol{\sigma}_{j}(k)$, e.g.
$\boldsymbol{\sigma}_{j}(k)\boldsymbol{\sigma}_{j}^{\dagger}(k)
=\mathbbm{1}$,
with $\boldsymbol{\sigma}_{j}(k)=\boldsymbol{\sigma}_{j}^{\dagger}(-k)$.
These conditions are equivalent, in matricial form, to
\begin{align}
  \label{eq:sigma_rel}
  \sigma_{j}^{[\e{j,l},\e{i,j}]}(k) =
    \left[\sigma_{j}^{[\e{i,j},\e{j,l}]}(-k)\right]^{*},\nonumber \\
  \sum_{i \in E_{j}}
  \sigma_{j}^{[\e{j,l},\e{i,j}]}(k)
  \left[\sigma_{j}^{[\e{j,m},\e{i,j}]}(k)\right]^{*} =
  \delta_{lm}, \\
  \sum_{i \in E_{j}}
  \sigma_{j}^{[\e{i,j},\e{j,l}]}(k)
  \left[\sigma_{j}^{[\e{i,j},\e{j,m}]}(k)\right]^{*} =
    \delta_{lm}\nonumber,
\end{align}
which can be seen as a generalization of one-dimensional scattering amplitudes
\cite{chadan}. For instance, for a node of degree 2, we can write 
\begin{eqnarray}
\boldsymbol{\sigma}(k)=\begin{pmatrix}
r^+(k) & t^{-}(k) \\
t^+(k) & r^{-}(k),
\end{pmatrix}
\end{eqnarray}
and the conditions $\boldsymbol{\sigma}(k)\boldsymbol{\sigma}(k)^\dagger=\boldsymbol{\sigma}(k)^\dagger \boldsymbol{\sigma}(k)=\boldsymbol{1}$, and $\boldsymbol{\sigma}(-k)=\boldsymbol{\sigma}(k)^\dagger$, reduce to 
\begin{eqnarray}
r^{+}(k)r^{-}(k)+t^{+}(k)r^{-}(k)&=&0\\
\end{eqnarray}

Referring to Fig. \ref{fig:transfref}, we must have at node $j$ and for the incoming wave $i\rightarrow j$, that
\begin{eqnarray}
    |\sigma_j^{\e{j,i},\e{i,j}}|^2+\sum_{p=1}^{d_j-1} |\sigma_j^{\e{j,p},\e{i,j}}|^2=1.
\end{eqnarray}

Let us now introduce the adjacency matrix technique for the evaluation of the Green's function, eqn. (\ref{eq:gfa}). The goal of this section is to state an alternative formulation based on the adjacency matrix $A_{ij}$ (not to be confused with $\mathcal A_j$), instead of defining it in terms of the scattering paths. Such a formulation is more compact and can be written, instead of having to enumerate all possible scattering paths, by the solution of a linear problem. 

The construction of the total scattering {amplitude} $T_{\Gamma}$ is equivalent to the following construction, based on the adjacency matrix. Given the adjacency matrix $A(\mathcal G)$, the Green's function between the incoming edge $e_i$ to the outgoing edge $e_f$, can be written as \cite{andrade4}
\begin{equation}
\label{eq:gfaa}
  G_{\Gamma,fi} = \frac{m}{i\hbar^{2} k}
  T_{\Gamma} e^{i k (x_{i}+x_{f})},
\end{equation}
where, given $\mathcal N_{i}$ the neighborhood of the node $i$ (the nodes at distance one) we have that the total transmission probability is given by
\begin{equation}
T_{\Gamma} = \sum_{j \in \mathcal N_{i}} \sigma_{i}^{[\{i,j\},e_{i}]} A_{ij} p_{ij},
\label{eq:transm}    
\end{equation}
where $i$ is the entry node and where the definition of $p_{ij}$ will be given in a moment. The quantity $\sigma_{i}^{[\{i,j\},e_{i}]}$ is the transmission probability incoming from the ``open" edge $e_i$, incoming to node $i$, and scattering into edge $j$. Similarly, we define  $\sigma_{n}^{[e_f,\{i,n\}]}$ as the transmission probability at the node $n$ from the edge $\{i,n\}$ into the ''open" edge $e_f$.

Let us note that in the expression above, the sum over the scattering paths has disappeared, and we are left with an expression that depends only on the transmission and reflection probabilities, which are equivalent to the boundary conditions, as we have seen. We will assume these as given for the time being.
The equivalence between the Green's function via the Gutzwiller formula relies upon the following assumptions 
 \cite{andrade4}:
\begin{enumerate}
\item At every vertex $j$ of the graph we define a scattering matrix
$\boldsymbol{\sigma}_{j}(k)$ associated with the boundary condition used
at the vertex $j$;
\item a particle that propagates along the edge ($i,j$) (if present), contributes a factor $z_{ij} = z_{ji} = e^{i k \ell_{ij}}$,
where $\ell_{ij}$ is assumed to be the metric distance between the nodes $i$ and $j$;
\item
at each existing edge, let it be ($i,j$), we introduce a factor $p_{ij}$, representing the probability flowing between $i$ and $j$, and $p_{ji}$, flowing in the opposite direction. The quantity $p_{ij}$ is interpreted as the weight associated with the flow of the particle on that edge. For each node $j$, and $\mathcal N_{j}^n$ as the neighborhood of the node $j$ minus the node $n$.
For the unitary equivalence between the scattering paths Green's function of eqn. (\ref{eq:gfa}) and the Green's function based on the adjacency matrix of eqn. (\ref{eq:gfaa}), the functions $p_{ij}$ must satisfy the equation:
\begin{equation}
  \label{eq:pij}
  p_{ij} =
    \sum_{l \in {\mathcal N_{j}^{n}}} z_{ij}
    \sigma_{j}^{[\e{j,l},\e{i,j}]}A_{jl}p_{jl}
    +\delta_{jn} z_{in}\sigma_{n}^{[e_f,\e{i,n}]},
\end{equation}
and similarly for $p_{ji}$, provided that we simply swap the indices $ij$. At each vertex $i$ we associate one $p_{ij}$ for every $j \in \mathcal N_{i}$. The last term in \eqref{eq:pij} is, as said above, the transmission amplitude at the
vertex $n$ from the edge $\e{i,n}$ to the lead $e_f$. By construction, the graph is simple, and the particle cannot hop on the node, thus $p_{ii}=0$ for all nodes.
\end{enumerate}
 Given the definition above of the $p_{ij}$'s, eqn. (\ref{eq:transm}) make sense: the total transmission probability is simply equal to the total flow from the incoming node, which must be equal to the flow to the exit node.

The remarkable property of the construction above is that \textit{infinite} sum over the scattering paths is replaced by a finite matrix inverse. 
This approach is powerful, as constructing the Green's function is equivalent to knowing everything about our quantum system \cite{economou}. For instance, bound-state energies can be obtained from the poles of the Green function, and the
 wave functions from the associated residues
\cite{andrade}.
In fact, $|T_{\Gamma}|^{2}$ represents the global transmission
probability from the lead $e_i$ to the lead $e_f$ and
it is constructed from the individual quantum amplitudes.
A similar construction was employed, before \cite{andrade4}, in \cite{ragoucy,caudrelier}.

\begin{figure}
    \centering
    \includegraphics[scale=.45]{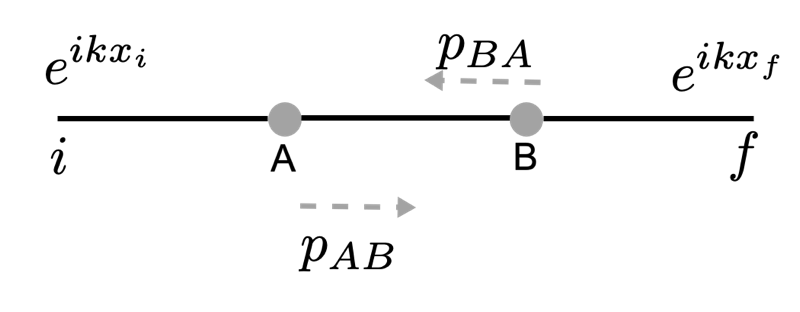}
    \caption{The graph above is the same scattering problem as in Fig. \ref{fig:scattering}, but the quantities $p_{AB}$ and $p_{BA}$ contain the information the graph's total Green's function. Their definition is in eqn. (\ref{eq:pij}). The transmission and reflection probabilities are the same as those of Fig. \ref{fig:scattering}.}
    \label{fig:andradeseverini}
\end{figure}

An example at this point might be useful to clarify how to use this construction. Consider for instance the graph in Fig. \ref{fig:scattering}, and the associated $p_{AB}$ and $p_{BA}$, as in Fig. \ref{fig:andradeseverini}. Because there is only one entry node of degree $2$, then we have $T_{\mathcal G}=t_A p_{AB}$, from eqn. (\ref{eq:transm}).
The equations for the $p$'s are then given by eqns. (\ref{eq:pij}):
\begin{eqnarray}
    p_{AB}&=& z_{AB}(r_B p_{BA}+t_B)\\
    p_{BA}&=&z_{BA} r_A p_{AB},
\end{eqnarray}
from which we obtain $$p_{AB}=\frac{{z_{AB}}t_B}{1-z_{AB}z_{BA} r_A r_B},$$
and finally, using $z_{AB}=z_{BA}=e^{ikl}$
\begin{eqnarray}
    T_{\Gamma}=t_A p_{AB}=\frac{t_A t_B {e^{ikl}}}{1-  r_Ar_B e^{{2}ikl}},
    \label{eq:twonodes}
\end{eqnarray}
which is the same expression as in eqn. (\ref{eq:transmfi}), and that we had obtained earlier from the sum over the infinite scattering path.

The construction presented so far is for general quantum graphs. But we need to supply the definitions for the transmission and reflection probabilities, which is where the Physics of the scattering is contained.

Clearly, we know from basic quantum mechanics that the $r$'s and the $t$'s cannot be independent, and so the entries of the matrix $\boldsymbol{\sigma}_j$ cannot either.
For instance, in a one-dimensional scattering, it is known that we must have $|r|^2+|t|^2=1$, following from the boundary conditions of eqns. (\ref{eq:k1}) and (\ref{eq:k2}). In the case of a star graph, if the wave is incoming from node $i$ to node $n$, the particle can be either reflected, or it can be transmitted in multiple directions, then we must have
\begin{eqnarray}
    |\sigma_n^{{i,n},{n,i}}|^2+\sum_{s\in \mathcal N^i_n}|\sigma_n^{{i,n},{n,s}}|^2=1,
\end{eqnarray}
or, written in terms of the elements of $\sigma's$
\begin{eqnarray}
    |r_n^{i,n}|^2+\sum_{s\in \mathcal N^i_n} | t_n^{i,s}|^2=1.
\end{eqnarray}
We can write one of these for each scattering process where the particle incomes from one of the nodes and scatters into the other neighbors, at the given node $n$, and these represent the constraints of our local scattering parameters.
The equations above clearly generalize the standard one-dimensional scattering on a heterogenous structure, like the one of a filament with multiple branches.

For naked graphs, in which there are no potentials on the edges, and the node is in a certain discrete sense isotropic, a set of quantum parameters conditions are given by
\begin{eqnarray}
    r_n&=&\frac{2}{d_j}-1\\
    t_j&=&\frac{2}{d_j},
    \label{eq:diffusive}
\end{eqnarray}
independently from the directionality. It is easy to see that these satisfy the conditions above, and physically represent the particle
diffusion isotropically at the node in all possible directions, with a probability only dependent on the local degree. For dressed graphs, e.g. a non-zero potential, this has to be generalized.

\subsubsection{Dressed graphs}
Insofar we have assumed that the propagation on the edge of the graph is free. We wish now to introduce a potential on the edges, e.g. if we have a local Hamiltonian of the form
\begin{equation}
  \label{eq:eigen}
  - \psi_{\e{i,j}}''(x) = k^2 \psi_{\e{i,j}}(x)+V_{\e{i,j}}(x) \psi_{\e{i,j}}(x),
\end{equation}
with a \textit{constant} potential $V_{\e{i,j}}=u_{ij}\equiv u$ along the edges and across the graph, the formalism of the previous section still applies, provided that we renormalize the edge length. This effective energy $u_{ij}$ represents an effective barrier in the propagation between sites.
First, let us briefly explain why a constant potential can act both as an approximation and why it can be physically motivated. 
First, a constant approximation of the potential is not uncommon in transport. In fact, the Kronig-Penney potential \cite{mermin76,Kittel87,kronigpenney}  is a common approximation in one-dimensional transport on a lattice. Since here our lattice is replaced by a generic graph, the assumption that the one-dimensional potential along each is the analog of a Kronig-Penney on a quantum graph, and can be seen merely as an approximation to estimate the dependence of the transport on such potential. The second comment is that a constant approximation can be solved exactly locally within the framework we introduced, and thus it is not only physically motivated but also of convenience. Nonetheless, we will then consider the general case first, and reduce to the constant case after.

As we have discussed earlier, in an arbitrary graph  we can always treat a vertex $j$ with its edges as a star
graph.
A star graph on $n$ vertices, $S_n$, is a graph where one
central vertex has degree $n-1$ and all other vertices have degree
$1$.
Consider a star graph as the one depicted in Fig. \ref{fig:fig1}
and let
$\Psi=\left(\psi_{\{j,1\}},\ldots,\psi_{\{j,n\}}\right)^{T}$.
The scattering solutions are given by
\begin{align}
  \label{eq:wavef2}
  \psi_{\e{i,j}}(x) = {} & e^{- i k_{ij} x}
                            + r_{j}^{[\e{j,i},\e{i,j}]} e^{i k_{ij} x},\\
  \psi_{\e{j,l}}(x) = {} & \sqrt{\frac{k_{ij}}{k_{jl}}}\,
                            t_{j}^{[\e{j,l},\e{i,j}]} e^{i k_{jl} x},
\end{align}
where
\begin{equation}
  k_{ij}=k_{ji}=\sqrt{\frac{2 m (E-u_{ij})}{\hbar^2}}=
  k\sqrt{1-\frac{u_{ij}}{E}},
\end{equation}
and $u_{ij}$ is the constant potential along the edge $\e{i,j}$.
Note that if $u_{ij}>E$, then $k_{ij}$ becomes complex and thus modes become evanescent. Thus, $u_{ij}$ acts as an effective cutoff for the energy of the particles that are transmitted on the graph.

Assuming the Neumann boundary condition
\begin{align}
  \label{eq:bc_delta}
  \psi_{\e{j,l}} = {} & \varphi_{j},
  \qquad \forall  l \in E_{j},
  \\
  \sum_{l \in E_j} \psi_{\e{j,l}}'
  = {} & 0,
\end{align}
we obtain the transmission and reflection amplitudes 
\begin{align}
  r_{j}^{[\e{j,i},\e{i,j}]}=  {}
  &
  \frac
  {k_{ij}-\sum_{l\in E_{j}^{i}} k_{jl}}{K},
    \nonumber \\
  t_{j}^{[\e{j,l},\e{i,j}]}= {}
  &
    \frac
    {2\sqrt{k_{ij} k_{jl}}}{K},
\end{align}
where $K=\sum_{l\in E_{j}} k_{jl}$, $E_{j}$ is the neighborhood of the
vertex $j$, and $E_{j}^{i}$ is the neighborhood of the
vertex $j$ but with the vertex $i$ excluded.

For a vertex with degree two
\begin{equation}
  \label{eq:R2e}
  r_{j}^{[\e{j,i},\e{i,j}]} =
  \frac{k_{ij}-k_{jl}}{k_{ij}+k_{jl}},
\end{equation}
\begin{equation}
  \label{eq:T2e}
  t_{j}^{[\e{j,l},\e{i,j}]} =
  \frac{2\sqrt{k_{ij} k_{jl}}}{k_{ij}+k_{jl}},
\end{equation}
which are the quantum amplitudes for a one-dimensional step potential.


The exact scattering Green's function for a dressed quantum graph
with adjacency matrix $A(\Gamma)$ can be written as
\begin{equation}
  G_{\Gamma} = \frac{m}{i\hbar^{2} k}
  T_{\Gamma} e^{i k_i x_{i}+i k_f x_{f}},
\end{equation}
where
$T_{\Gamma} = \sum_{j \in E_{i}} \sigma_{i}^{[\e{i,j},e_{i}]} A_{ij} p_{ij}$,
and
\begin{equation}
  \label{eq:pij2}
  p_{ij} =
    \sum_{l \in {E_{j}^{n}}} z_{ij}
    \sigma_{j}^{[\e{j,l},\e{i,j}]}A_{jl}p_{jl}
    +\delta_{jn} z_{in}\sigma_{n}^{[e_f,\e{i,n}]},
\end{equation}
with
\begin{equation}
  z_{ij} = e^{i k_{ij} \ell_{ij}} = e^{i k\sqrt{(1-u_{ij}/E_k)}\ell_{ij}}.
\end{equation}

Let us now assume that $u_{ij}=u$ and $l_{ij}=l$.  Then, we can replace $l\equiv l(k)=l\sqrt{1-\frac{u}{E_k}}$ on every edge of the graph.

If all the wave vectors are equal, i.e, $k_{ij}=k$, we have
  \begin{eqnarray}
    \label{eq:R2c}
    r_{j} &=& \frac{2}{d_j}-1,\\
    \label{eq:T2ec}
    t_{j} &=& \frac{2}{d_j},
  \end{eqnarray}
which are the well-known quantum amplitudes for the Neumann boundary
condition for ``naked'' quantum graphs we discussed earlier. As such, we can use the same framework as before, provided that we renormalize the length at each particular value of $k$, which explains why these types of graphs are  ``dressed".
Thus, in the dressed graph case, we will have an extra parameter to analyze, which is $u$.

\section{Classical model of filament growth and effective quantum graph}
At this point, we need to discuss how the filaments are formed in our model. Below, we provide the simplest non-trivial model which incorporates the voltage for the filament growth. 

First, we consider a certain narrow atomic region, of size $W\times W\times H$, where $H$ is the horizontal line dimension of the region, and $H$ is its vertical dimension, in atomic units. Let us now assume that the top (anode) and bottom (cathode) of the region $z=H$ and $z=1$, are connected to a certain battery applying a voltage $V$. We assume then that the anode and cathode have a certain voltage $v_a=\frac{V(t)}{2}$ and $v_c=-\frac{V(t)}{2}$, and with an opposite sign if the battery's voltage changes sign. These are the simplest boundary conditions. We follow the following algorithm for every time $t$

\begin{itemize}
\item We assume that at the anode and cathode, we have a certain time-dependent voltage $V(t)$ applied, driven by the battery. Inside the region, the vector field is given by $\vec E(x)=-\vec \nabla V(p)$, where $V(p)$ is the potential inside the region. To calculate the potential $V(p)$, we solve for the equation
\begin{eqnarray}
    \triangle V_t(\vec x)&=&0\\
    V|_{\mathcal B_a(t)}&=&\frac{V(t)}{2}\\
    V|_{\mathcal B_c(t)}&=&-\frac{V(t)}{2}
    \label{eq:laplacet}
\end{eqnarray}
where ${\mathcal B_a}$ is the subset of points associated with the anode, while ${\mathcal B_c}$ is associated with the cathode and $\triangle=\vec \nabla \cdot \vec \nabla$. The equation above can be solved numerically using an iterative method. Once we have obtained the solution $V(\vec x)$ everywhere, $\vec E(x)$ can be obtained.

\item Insofar, however, we have not discussed how the filament is formed dynamically. We assume that if $V>0$, a certain density per unit of time and unit of voltage $\rho_0$ is associated to the creation of particles of mass $m$,  formed at the center of the cathode, and if $V<0$ it is formed at the center of the anode, with zero momentum. Then, we solve numerically for the equation of motion for each particle $i$, which are assumed to be  all identical ions of charge $e_c$ abd $m_p$,
\begin{eqnarray}
    \frac{d^2 \vec x_i}{dt^2}=- \frac{e_c}{m_p} \vec E(\vec x)+W(t).
    \label{eq:mdd}
\end{eqnarray}
It is easy to see from eqn. (\ref{eq:mdd}) that we are using a molecular dynamics approach to the construction of the filament between the two leads.
If a particle reaches the anode, or cathode, or touches another particle, a filament is formed. If the filament touches the cathode, then the number of particles is associated with $\mathcal B_c(t)$, and if at the anode they are associated with $\mathcal B_a(t)$. Following this prescription, the electric field $\vec E_t$ is also dynamic, but the time label is only to capture the electric field at a particular time $t$ in the simulation. Because the speed of the particles is assumed to be much lower than the speed of light ($V$ small enough), $p_i/m\ll c$, we are still solving for electrostatic equations.
The quantity $W(t)$ is a Wiener noise associated with the random formation of the filament. Because of this, filament formation is a stochastic process. Particles can also tick to the partly formed filament: if a certain particle is directly in the vicinity of the filament, the particle can stick and grow the filament. Thus, in principle, the filament can grow both in horizontal and vertical directions.

\item If the electrodes are connected by a sequence of particles, we construct the connected component of the effective graph $\mathcal G_(t)$, which encodes the adjacency of the atoms. We construct the set of vertices $\mathcal N_c$ and $\mathcal N_a$ which are the vertices connected to the cathode and anode respectively via a depth-first-search algorithm. These are the vertices which associated with the Lippman-Schwinger states.

\item If the voltage flips sign, then the graph $\mathcal G(t)$ can decay. We associate a certain probability $p_{d}$ per unit of time such that the vertices (the atoms) can be removed. In parallel, the growth process can restart from the other lead, and particles appear at random at the other end. The filament formation then restarts. As such, the device we are simulating is apolar.
\end{itemize}

\begin{figure*}
\includegraphics[width=\linewidth]{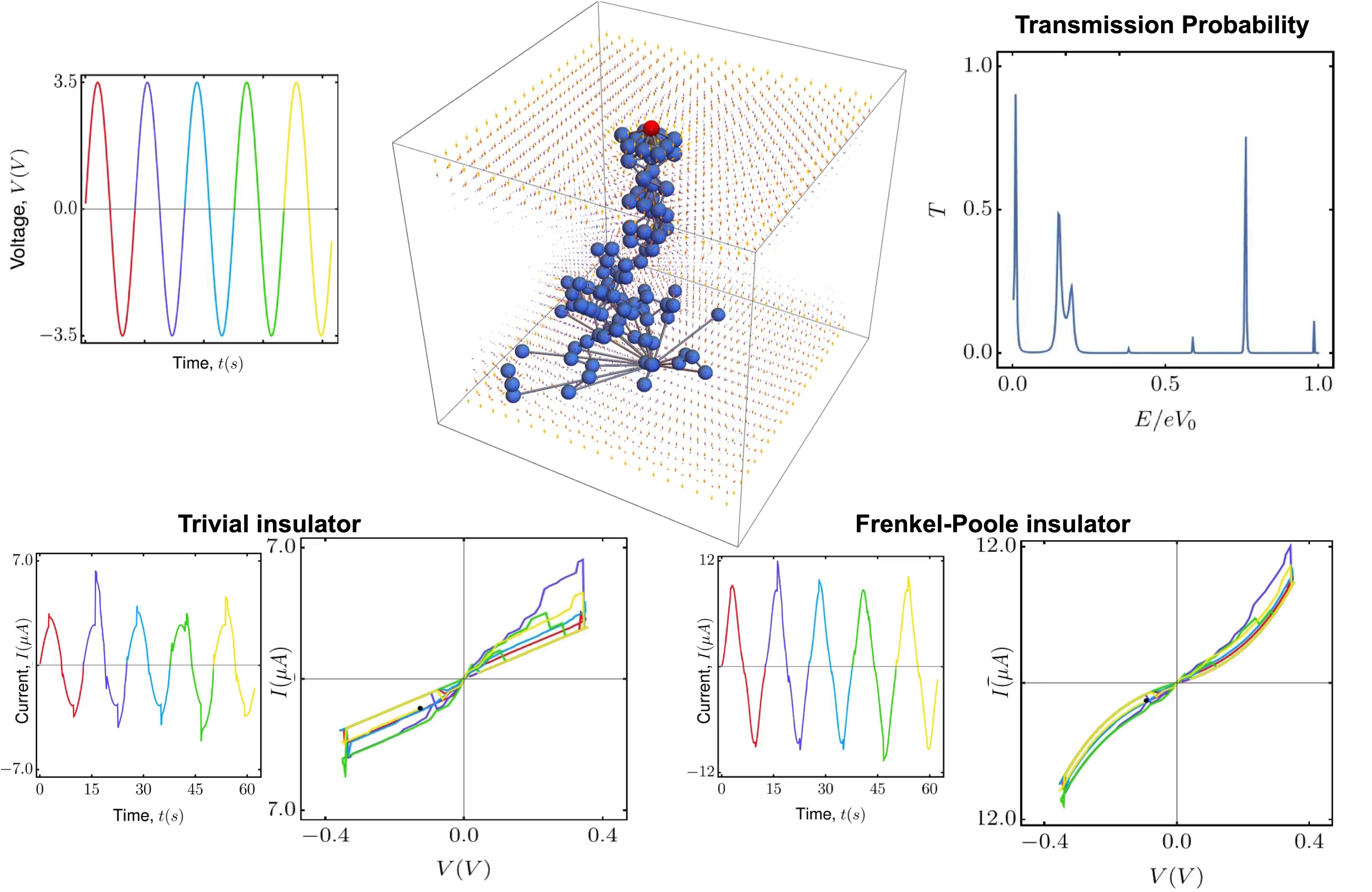}
\caption{In the center images, we show a filament formed during an aggregation process, due to the ion migration,  induced by the electric field, and equivalent to a quantum graph. We see the transmission spectrum on the top right panel. The voltage is shown on the left panels. In the central figure, the size of the atoms has been reduced to a point to show the connectivity graph. For this figure, we have chosen an insulator with constant conductance. At the bottom of the graph in the center panel, all the points on the surface are connected to a single node as in Fig. \ref{fig:lippsch}. Five cycles of filament formation (color-coded). In the  bottom panels, we see the final cycle filament formed. In the corresponding I-V curves, the color-coded hysteresis curves. The corresponding currents are shown on the left of each I-V curve panel. The difference between the left and right is the choice of the insulator. In the case of the right, we have chosen a Poole-Frenkel insulator, with exponential leakage. The Schottky barrier is approximately $V_0= 0.3 V$. This implies an exponential envelope, usually observed in experiments \cite{halbritter}. In the left panel, we have chosen an insulator constant with respect to the voltage. The transmission probabilities as a function of the energy are shown in the bottom right of each filament.}
\label{fig:fil}
\end{figure*}

An example of the filament formed via this process is shown in Fig. \ref{fig:fil}, with the corresponding current and voltage cycle, the I-V diagram and the transmission probability as a function of the energy for the final junction. 

An important comment is that at this point, the device's effective conductivity can be calculated from the Landauer formula,
using $\sigma_r=\frac{I_t}{V_t}$, where $I_t$ is the effective.
However, in experimental realizations, it is hardly possible to isolate the filament uniquely. As a result, the effective conductivity of the device is typically the one of the filament plus a parallel conductive device, whether this is a dielectric or an actual resistor. It is often argued in fact that even at a few Kelvins, it is not possible to isolate the linear resistive switching effect, but one has to consider other types of tunneling effects \cite{halbritter}. For this reason, we add to the resistivity an effective conductivity, mimicking a classical or semiclassical resistor. The effective current is then given by
\begin{eqnarray}
	I(V) &=& I_{ L}(V) + I_{R}(V)  \nonumber \\
	&=&
	G_0 \int_{-\infty}^\infty dE (f_L(E)/N-f_R(E)/M)\cdot \nonumber\\
 &&\hspace{2cm}\cdot(\frac{h}{2e^2}\sigma_0(V)+ T(E))
	\label{eq:totalCurrentC2}.\nonumber \\
\end{eqnarray}
For this paper, we use both a classical resistor and thus $\sigma_0$ can either be a constant or a voltage-dependent quantity. In the case it is a constant, then we will observe at low temperatures just a resistive switching, without nonlinearity, but even if the device's filament is not present, a current. If $\sigma_0(V)$ is nonlinear, then the observed device's current can be nonlinear. In this paper we also consider a tunneling effect \cite{frenkelpoole,simmons}, with $\sigma_0(V)\sim \sigma_0 e^{V/V_c}$, with $V_c\approx 0.15$ volts, with $\frac{h}{2e^2}\sigma_0\sim 10^{3} \Omega^{-1}$. This is the number of the same order of magnitude as the one typically observed in tunneling experiments, and such that at voltages of the order $V\sim 1$ volts, we obtain currents of the order of the microamperes.

\begin{figure*}[b!]
    \centering
    \includegraphics[width=\linewidth]{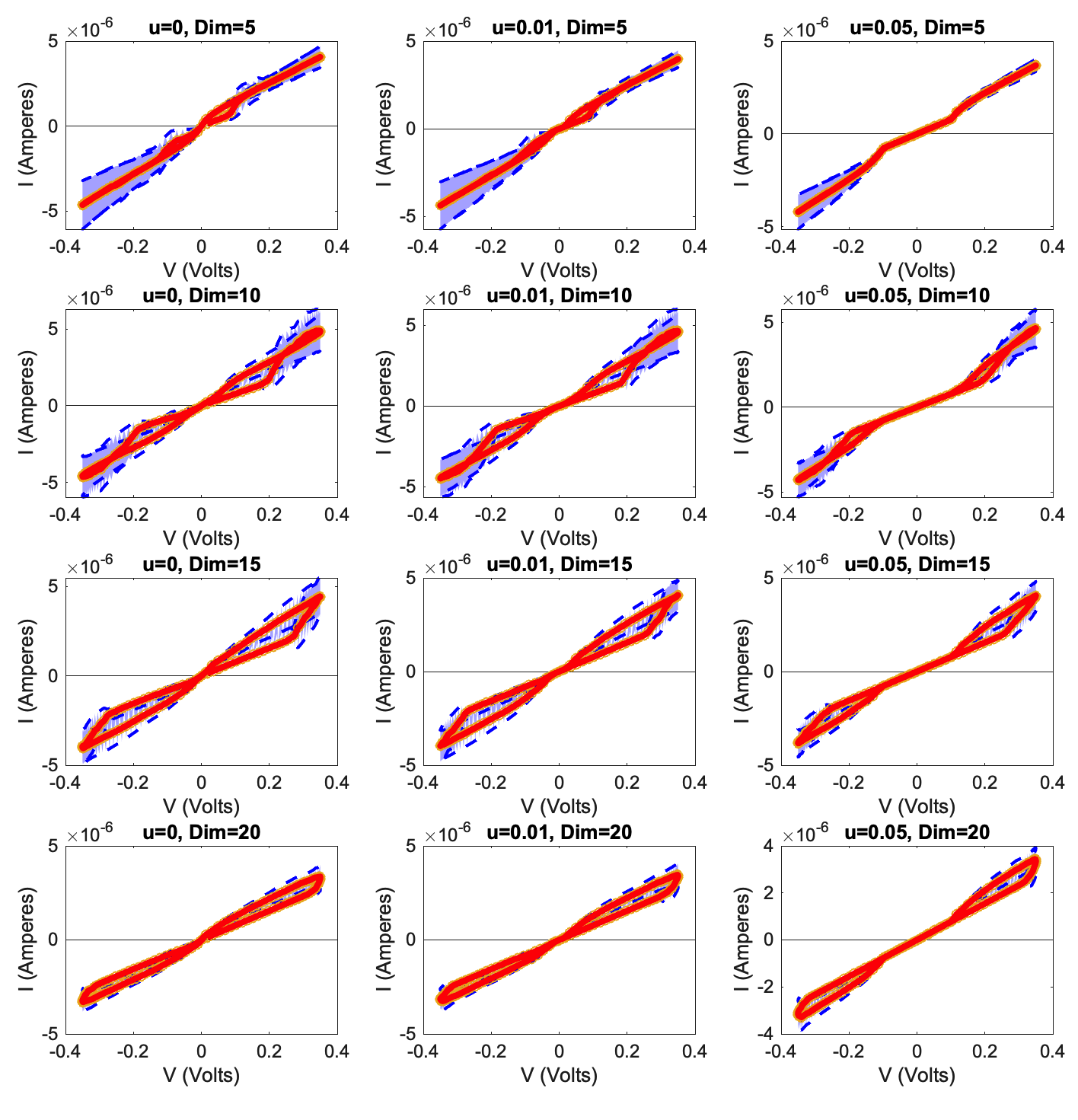}
    \caption{ Current-Voltage characteristic within the context of the model developed in this paper averaged over 100 Monte Carlo samples, and as a function of the dressing parameter $u$, for $u=0$, $u=0.01$ and $u=0.05$ and $x=0$.}
    \label{fig:dressed}
\end{figure*}
\section{Numerical simulations}
We now discuss the results of our numerical experiments.
The goal is to provide some qualitative $I-V$ curves and the dependence on some key underlying parameters.

\subsection{Parameters choice} In our simulations, the parameters have been chosen as follows. First, the depth of the region is such that approximately $10$ atoms can be in a straight line, a number compatible with the approximations we wished to make from the start (e.g. a discrete rather than continuous filament). In particular, we have chosen an arbitrary time unit fixed, however, by the following prescription. In the molecular dynamics of eqn. (\ref{eq:mdd}) the parameters have been chosen such that within $10-50$ time units, we observe a filament formation at a certain target frequency $\omega^*$, assuming $V(t)=V_0 \cos(\omega^* t)$, with $V_0=0.35$ V. In terms of the voltage drive, the time units can be expressed such that under the sinusoidal drive at $V\approx 0.3$ we do observe a switching, which is a physically plausible number for the switching.


Also, for simplicity, we have assumed that the Fermi chemical energy $\mu_F$ is much smaller than the voltage applied, $e V$. In doing so, we have essentially constrained our parameters to be at the typical frequency at which, depending on the ions that form the filament, the observation of the $I-V$ hysteresis first shows a memristive effect. At this point, the free parameters are then the strength of the stochastic force in eqn. (\ref{eq:mdd}) and the potential $u$ in the quantum graph. The strength of the stochastic force has been chosen to be $\langle W(t)^2 \rangle_t = \sqrt{0.005}$, which was the smallest value such that at every single simulation we could obtain a different filament. This is dictated by the necessity that we wish to perform a Monte Carlo analysis of the $I-V$ curves. The molecular dynamics simulations were instead performed integrating the Newtonian equations of motion for the particles in the electric field in the narrow region.

\subsection{Analysis of the results}
Provided that our model of filament growth is limited in many ways, let us first comment that we had first written a 2-dimensional code. In spirit, the results are similar to those that we have obtained here, but some comments are in order. In the three-dimensional case we discussed here, the filament attracts particles from three dimensions, meaning that it is much more likely for a shortcut to be found. This is why we found that there are many more ways to have significant jumps in current in three dimensions rather than two. Moreover, we have found that a good model for the growth of the filament in two dimensions, as an alternative for the voltage-induced molecular dynamics, was a diffusion-limited aggregation process \cite{pietronero}, without having to solve for the Laplace equation. These results will be discussed elsewhere, but from the point of view of the present paper, it is experimentally more relevant to study the three-dimensional case.

While this paper aims to provide a qualitative agreement between previously obtained experimental results and theoretical ones, some comments are in order.
For that purpose, let us look at Fig. \ref{fig:fil}, where several voltage cycles were applied to the device with $u=0$. We see the filament formed on the negative voltage side at the center.\footnote{The video of the filament formation for  Panel \ref{fig:fil} is shown at 
\url{https://youtu.be/mhgoE0MWGk0}.}

We have found that in the vast majority of our numerical experiments, only one filament was formed. The reason is that as the filament is formed, it attracts any free particle in the vicinity as a result of eqn. (\ref{eq:laplacet}). This implies that although the filament has a finite thickness, the charge propagation is quasi-one-dimensional. However, gaps in the filaments can be filled by floating particles, leading to discrete jumps in the resistivity. This is the jump we see for instance in Fig. \ref{fig:fil} (bottom left IV). The (middle right) is the voltage profile applied to the device.  The vertical jumps, both on the positive and negative side, due to the resistive switching, e.g. the filament reaching the cathode/anode. The curves in Fig. \ref{fig:fil} represent several realizations of the filament. The averaged curves are shown in Fig. \ref{fig:dressed} for various sizes of the junction and values of the parameter $u$. 

We have also performed a slightly more refined statistical analysis of the conductance distributions. In particular, in Fig. \ref{fig:dressed} we plot the averaged hysteresis curve over 100 realizations of the filament, and for three values of $u=0,0.01,0.05$ and junction size of $5, 10,15,20$ atoms. The shadow region represents the error across various realizations of the filament. We see that for larger values of the gap, the memory effect is more noticeable, due to the intuitive fact that there are more possibilities for the filament to be formed. In Fig. \ref{fig:fil} (bottom) we show different cycles of hysteresis for two types of insulators.\footnote{A video for the Frenkel-Poole insulator is 
\url{https://youtu.be/uUrPZnSrIpQ}
. For the insulator model which is constant in the voltage, a video is available at 
\url{https://youtu.be/TfrhEuasAD4}
.}

\begin{figure*}
    \centering
    \includegraphics[width=\linewidth]{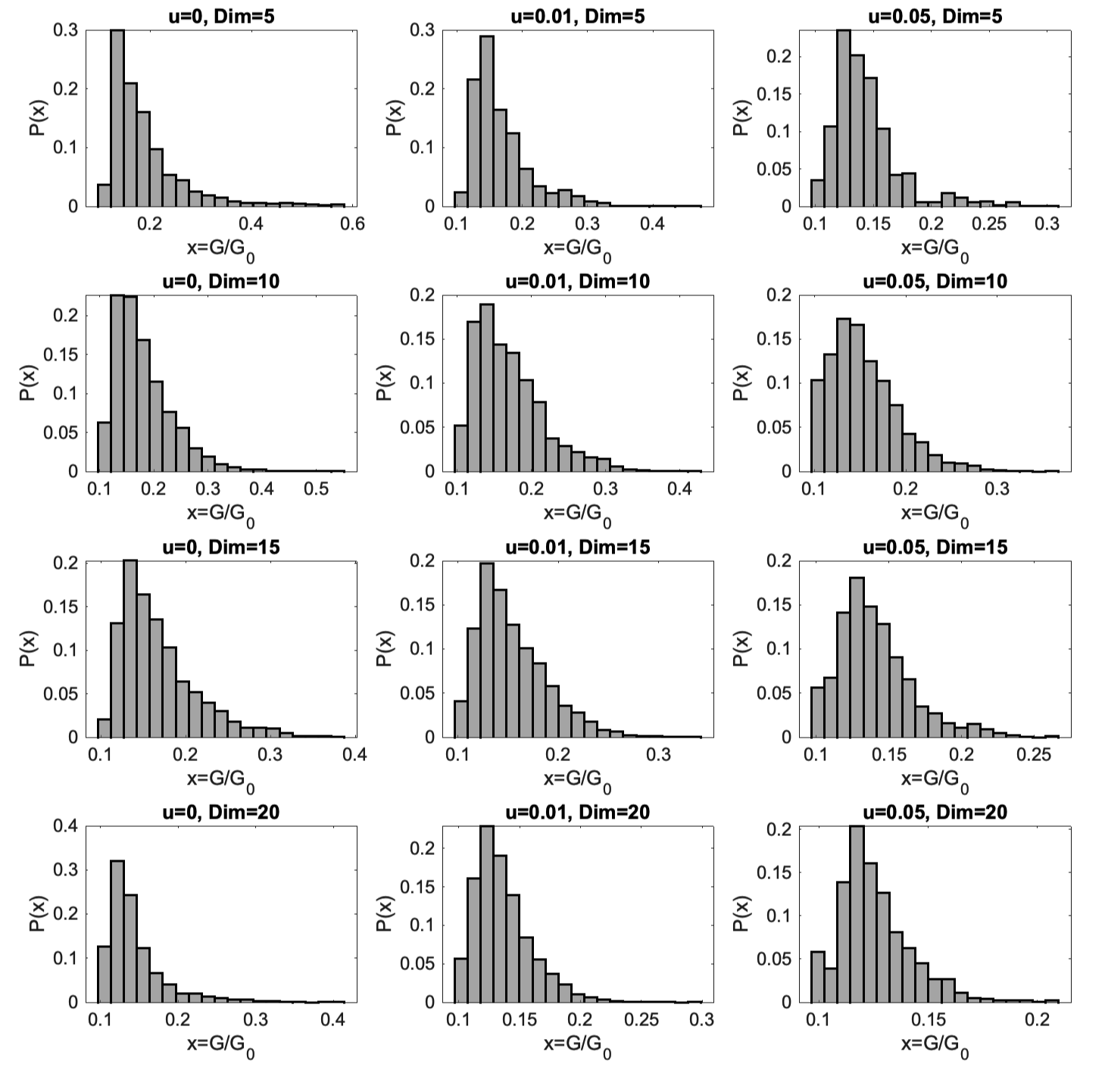}
    \caption{Conductance distribution as a function of the gap size (in atoms, Dim) and the parameter $u$. For smaller regions, we see that the probability of having a conductance near one is non-zero. For larger gaps, the probability decreases and peaks near smaller values. The parameter $u$ broadens the distribution. Each figure is averaged over 100 realizations of the filament.}
    \label{fig:gapsize}
\end{figure*}

We then note that, for $u=0.01$, the switching to the ON state is to a lower resistive state. In particular, because of this, the tunneling phenomenon is such that the switching to the ON is almost hidden in the nonlinearity caused by the tunneling. This feature was also observed in the measurements of niobium oxides \cite{halbritter}. Thus, this qualitative behavior of the pinched hysteresis obtained by this model is realistic.  The distribution of the conductance is plotted in Fig. \ref{fig:gapsize}.  Overall, this provides a hint about the relationship between the size of the gap and the conductance (see for instance \cite{beenaker}, Fig. 19), expected from an Anderson model with finite localization length \cite{pichard}. An important point to make is that the parameter $u$, in particular for larger values, can be interpreted in terms of an (effective) Schottky barrier to electron propagation.

\section{Conclusions}
The present paper introduced a framework to study quantum transport within the context of filamentary switching. The method we developed is not based on the quantum scattering in a continuum or a lattice approximation but assumes that the filament is finite-dimensional and discrete, e.g. charge carriers can move on a discrete graph.  This paper is based on the Landauer-B\"{u}ttiker formula applied to quantum graphs, which provides the theory to study one body scattering on discrete structures such as metric graphs. In addition, we have used a model for the filament growth which is based on a molecular dynamics simulation. The molecular dynamics we used is the simplest model that supports filament growth and is based on the solution of the Laplace equation and the derivation of the electrical vector field in the narrow region.

Yet, although our results are not compatible with previous numerical studies for a continuous saddle for silver ions \cite{gubicza1}, and the experimental results for niobium pentoxide \cite{halbritter}, in which filamentary switching occurs, they are more similar to the junction model by Milano and collaborators \cite{milano1,milano2, carmilano}, for which stochastic models are currently being developed \cite{stochast1,stochast2}. In addition, the distribution of conductances in Fig. \ref{fig:gapsize} qualitatively is consistent with the conductance distributions arising from random matrix theory \cite{beenaker}.

As far as we understand this is the first time that a non-many body quantum approach has been used for a realistic dynamical filament formation and where we model the discrete nature of the filament, rather than it being assumed continuous. In this sense, we believe our paper provides a new methodology to analyze quantum transport on these structures. The advantage is in the scalability of our numerical codes: our codes scale as $\mathcal O(E^3)$ (which comes from the matrix inversion for evaluating $p_{ij}$), where $E$ is the number of active channels, or edges of the graph (between atoms/molecules) in the narrow region. For the case with ballistic motion, the code scales as $\mathcal O(d_{max} N E^3) $, but still subexponentially, where $N$ is the number of atoms and $d_{max}$ the maximum graph degree. As we discussed in the text, this extra factor comes from the construction of the scattering amplitudes $\sigma_j$'s.
This paper did not consider a particular type of material, but we focused on generic features of the I-V curves which are phenomenological in nature.
While this is just a first step in this direction, various improvements can be made. First of all, a more precise model of the filament formation can be introduced, which carefully
 models the evolution of the filament over time. Moreover, it allows the introduction of a variety of modeling assumptions about the scattering and channel properties of the filament and atoms. One important simplifying assumption used here was that all the graph edges were of equal length $l$, the unit of distance between atoms. Relaxing such assumption allows also to have localization phenomena and drastic reductions of the conductance (induced by small $T_{fi}$ between the leads) due to chaotic scattering in the graph \cite{qgchaos}.

One drawback of our methodology is that the evaluation of the scattering amplitudes scales with the number of edges of the graph, the reason for which we needed to focus on a few atoms. However the calculation of the scattering amplitude can be scaled, and study larger graphs and larger regions. In fact, the technique scales most quadratically in the number of nodes or atoms, and while challenging, it is feasible using a proper computing architecture. Also, because the method requires the inversion of a large matrix, it is also possible to use GPUs for this purpose. 

 Nonetheless, some of the findings of this paper are particularly interesting. For instance, the non-monotonicity of the ON state on the dressing parameter is an interesting feature that can be studied experimentally.  Moreover, this approach can be extended also beyond the one-body Hamiltonian, which can be generalized to a local one-dimensional quantum field \cite{mintchev1,mintchev2,mintchev3,mintchev4}.

 These and other generalizations and improvements will be considered in future papers.

 \begin{acknowledgements}
 This work was carried out under the auspices of the NNSA of the U.S. DoE at LANL under Contract No. DE-AC52-06NA25396. FC was also financed via an LDRD IMS-Rapid Response grant, and this research was supported in part by grant NSF PHY-1748958 to the Kavli Institute for Theoretical Physics (KITP).
The work of F.M.A. was partially supported by Co\-or\-dena\c{c}\~{a}o de
Aperfei\c{c}oamento de Pessoal de N\'{i}vel Superior (CAPES, Finance
Code 001). It was also supported by Conselho
Nacional de Desenvolvimento Cient\'ifico e Te\-cnol\'ogico (CNPq), Instituto
Nacional de Ci\^{e}ncia e Tecnologia de Informa\c{c}\~{a}o Qu\^{a}ntica
(INCT-IQ). FMA acknowledges financial support by
CNPq Grant 313124/2023-0. The authors are indebted to András Halbritter for several comments on the manuscript.\\ 
 \textbf{Author contributions}. F. Caravelli conceived the study. F. Andrade, F. Caravelli and A. A. Silva contributed to the theoretical analysis in this order. F. Caravelli and A. A. Silva wrote the code for the analysis.  F. Caravelli wrote the first draft. All authors contributed to the writing of the draft.\ \\
 \textbf{Data availability.} A Mathematica code has been assigned a FigShare identifier \cite{figshare}.
 \end{acknowledgements}


\end{document}